%% file: SnPRL-v19-APZ.tex
\RequirePackage{lineno}
\documentclass[aps,prl,twocolumn,showpacs,superscriptaddress,groupedaddress]{revtex4}
\usepackage{graphicx}  
\usepackage{dcolumn}   
\usepackage{bm}        
\usepackage{amssymb}   
\usepackage{xcolor}
\usepackage{natbib}
\usepackage{multirow}

\newcommand\tstrut{\rule{0pt}{2.4ex}}
\newcommand\bstrut{\rule[-1.0ex]{0pt}{0pt}}

\newcommand{\A}[1]{$^{#1}$}
\newcommand{\vlk}{V_{\text{low-}k}}

\begin{document}



\title{Pairing-quadrupole interplay in the neutron-deficient tin nuclei: \\first lifetime measurements of low-lying states in $^{106,108}$Sn} 
\input PRL_AuthorList.tex    

\begin{abstract}
The lifetimes of the low-lying excited states $2^+$ and $4^+$ have been directly measured in the neutron-deficient $^{106,108}$Sn isotopes. 
The nuclei were populated via a deep-inelastic reaction and the lifetime measurement was performed employing a differential plunger device. 
The emitted $\gamma$ rays were detected by the AGATA array, while the reaction products were uniquely identified by the VAMOS++ magnetic spectrometer. 
Large-Scale Shell-Model calculations with realistic forces indicate that, independently of the pairing content of the interaction, the quadrupole force is dominant in the $B(E2; 2_1^+ \to 0_{g.s.}^+)$ values and it describes well the experimental pattern for $^{104-114}$Sn; the $B(E2; 4_1^+ \to 2_1^+)$ values, measured here for the first time, depend critically on a delicate pairing-quadrupole balance, disclosed by the very precise results in $^{108}$Sn. 
This result provides insight in the hitherto unexplained $B(E2; 4_1^+ \to 2_1^+)/B(E2; 2_1^+ \to 0_{g.s.}^+) < 1$ anomaly. 
\end{abstract}

\maketitle

A little over a decade ago, the Sn isotopes were considered the paradigms of pairing dominance: low-lying states of good seniority, nearly constant $J^\pi = 2_1^+$ excitation energies and parabolic $B(E2; 2_1^+ \to 0_{g.s.}^+)$ behavior. 
The latter was observed for $A \geq 116$.
For the lighter species, experimental results on transition probabilities were scarce as the presence of low-lying isomeric states hindered direct measurements of lifetimes below them. 
From Coulomb-excitation measurements with radioactive ion beams only the reduced transition probability between the first excited $2^+$ state and the ground state could be determined~\cite{banu2005sn, vaman2007sn, cederkall2007sub, ekstrom2008sn, kumar2010enhanced, bader2013quadrupole, doornenbal2014intermediate, kumar2017noevidence}. 
Within experimental uncertainties, they suggest a rather-constant behavior for $106 \leq A \leq 110$, instead of the parabolic trend expected when isovector $T=1$ pairing dominates. 
This ``plateau'' includes also the stable $^{112,114}$Sn nuclei, for which measurements of both $B(E2; 2_1^+ \rightarrow 0_{g.s.}^+)$ and $B(E2; 4_1^+ \rightarrow 2_1^+)$ values exist~\cite{jonsson1981collective}. 
Before this work, the $B(E2; 4_1^+ \to 2_1^+)$ values were completely absent in the neutron-deficient Sn isotopes.

The experiment described in this Letter was devoted to determine the strength of $2_1^+ \to 0_{g.s.}^+$ and $4_1^+ \to 2_1^+$ transitions in $^{106,108}$Sn by measuring the lifetime of $2_1^+$ and $4_1^+$ states. 
Although several theoretical interpretations have been proposed~\cite{ansari2005study, banu2005sn, ekstrom2008sn, Maheshwari2016asymmetric, PhysRevLett.121.062501}, the evolution of the $B(E2;2_1^+ \to 0_{g.s.}^+)$ values in the light Sn nuclei remains puzzling.
The experimental and theoretical results that will be presented in this Letter provide a further insight, which reveals the underlying structure of the light tin isotopes, namely the counterbalance of quadrupole and pairing forces in the Sn isotopic chain. 

A multi-nucleon transfer reaction, that is commonly used to investigate neutron-rich nuclei~\cite{szilner2007multinucleon, corradi2009multinucleon, valiente2016gamma}, was unconventionally adopted to populate the Sn isotopes close to the proton drip line, so the isotopes of interest were populated in the collision of a $^{106}$Cd beam and a $^{92}$Mo target. 
The beam-target combination and beam energy were selected as a compromise between two requirements.
On one hand the reaction fragments energy had to be sufficiently high to allow their identification by the spectrometer.
On the other, in order to perform the lifetime measurement, the population of the states above the $6^+$ isomers had to be minimized; this condition imposed an upper limit on the excitation energy and consequently on beam energy, even at the expense of the cross section to populate more exotic species.
The $^{106}$Cd beam, provided by the separated-sector cyclotron of the GANIL facility at an energy of 770 MeV, impinged onto a 0.7 mg/cm$^2$ $^{92}$Mo target. 
The lifetime measurement was performed with the Recoil Distance Doppler-Shift (RDDS) method~\cite{dewald1989differential, dewald2012developing, valiente2009lifetime}. 
The target was mounted on the differential Cologne plunger with a 1.6 mg/cm$^2$ thick $^{nat}$Mg degrader down stream. 
In order to measure the lifetimes of interest, 8 different target-degrader distances in the range 31-521 $\mu$m were used. 
The complete (A,Z) identification, together with the velocity vector for the projectile-like products was obtained on an event-by-event basis using the VAMOS++ spectrometer \cite{pullanhiotan2008improvement, rejmund2011performance, vandebrouck2016dual}, placed at the grazing angle $\theta_{lab}$=25$^\circ$. 
In coincidence with the magnetic spectrometer, the $\gamma$ rays were detected by the $\gamma$-ray tracking detector array AGATA~\cite{akkoyun2012agata, clement2017conceptual}, consisting of 8 triple-cluster detectors placed at backward angles in a compact configuration (18.5 cm from the target). 
The combination of the pulse-shape analysis \cite{bruyneel2013correction} and the Orsay Forward-Tracking (OFT) algorithm \cite{lopez2004gamma} allowed to reconstruct the trajectory of the $\gamma$ rays emitted by the fragments. 
More details about the ion identification and the analysis procedure can be found in Refs.~\cite{siciliano2017appb, siciliano2017ncc}.

\begin{figure*}
\centering
\includegraphics[width=\textwidth]{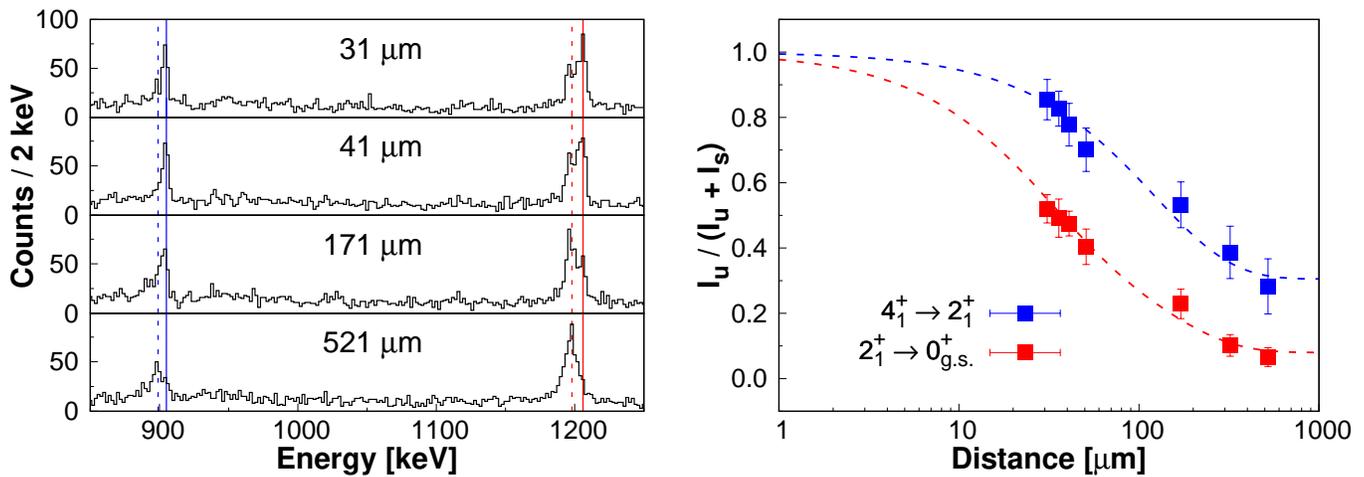}
\vspace{-3mm}
\caption{\label{fig:108Sn_spectra} (Color online) (left) Doppler-corrected $\gamma$-ray energy spectra of $^{108}$Sn for different target-to-degrader distances, gated on the Total Kinetic-Energy Loss (TKEL). For each transition the \textit{unshifted} ($u$) and \textit{shifted} ($s$) centroids are indicated by a solid and a dashed line, respectively. (right) Ratio of the transition components intensity as a function of the distance, obtained by gating on the TKEL. The dashed lines represent the fitted decay curves for the two excited states.}
\end{figure*}

\begin{figure}[t]
\includegraphics[width=0.48\textwidth]{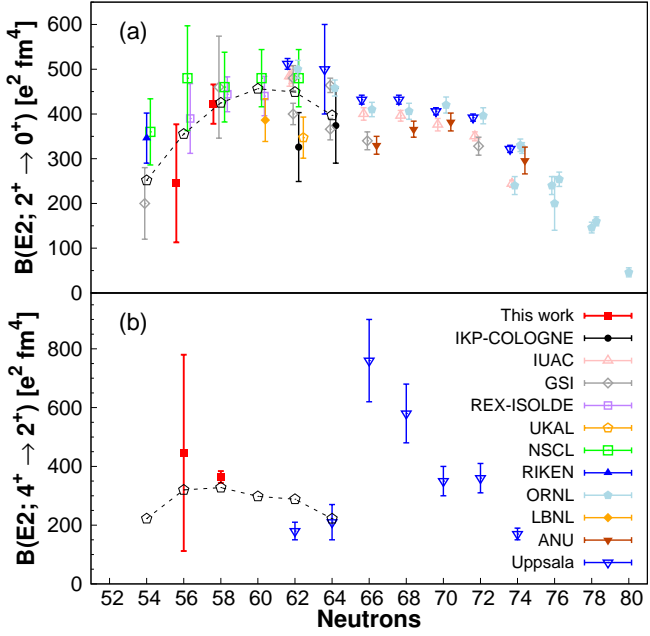}
\vspace{-3mm}
\caption{\label{fig:B(E2)} (Color online) Reduced transition probability $B(E2)$ for the (a) $2_1^+ \rightarrow 0_{g.s.}^+$ and (b) $4_1^+ \rightarrow 2_1^+$ transitions along the Sn isotopic chain. The results of the present work (red squares) are compared with those from previous experiments~\cite{jonsson1981collective, radford2004nuclear, banu2005sn, vaman2007sn, cederkall2007sub, orce20072, ekstrom2008sn, doornenbal2008enhanced, kumar2010enhanced, allmond2011coulomb, jungclaus2011evidence, kumbartzki2012transient, bader2013quadrupole, guastalla2013coulomb, doornenbal2014intermediate, allmond2015investigation, kumbartzki2016z, kumar2017noevidence, spieker2018shape}. The predictions from Large-Scale Shell-Model calculations from the I.3.4 interaction are also shown (open black pentagons).}
\end{figure}

Thanks to the precise determination of the ion velocity vector and the identification of the first interaction point of each $\gamma$ ray inside AGATA, Doppler correction was applied on an event-by-event basis. 
The magnetic spectrometer directly measured the fragments velocity after the degrader ($\beta_{after} \approx 9\%$). 
However, for each $\gamma$-ray transition two peaks were observed, related to its emission before and after the Mg foil: the $\gamma$ rays emitted after the degrader are properly Doppler corrected, while those emitted before are shifted to lower energies because of the different velocity of the reaction fragment ($\beta_{before} \approx 10\%$). 
The relative intensities of the peaks area as a function of the target-degrader distance are related to the lifetime of the state of interest.  
Figure~\ref{fig:108Sn_spectra} (left pad) shows the Doppler-corrected $\gamma$-ray energy spectra of $^{108}$Sn for several distances, where a Total Kinetic Energy Loss (TKEL) gate has been imposed in order to reduce the direct feeding of the higher-lying states~\cite{valiente2009lifetime, mengoni2009lifetime}. 
For this nucleus the energies of $8_1^+ \to 6_1^+$ and $2_1^+ \to 0_{g.s.}$ transitions are similar, so traditional methods cannot be used to measure the lifetime of the $2_1^+$ state. 
Thus, the TKEL$\le 21$ MeV condition was imposed until the $8_1^+ \to 6_1^+$ transition peaks became negligible and the measured lifetime of both $4_1^+$ and $2_1^+$ states remained constant, even for more restrictive conditions; such a procedure allowed us to take into account just the $6_1^+$, $4_1^+$ and $2_1^+$ states in the measurement of the lifetimes via Decay-Curve Method (DCM).
In Fig.~\ref{fig:108Sn_spectra} (right pad) the decay curves are presented for the $4_1^+$ and $2_1^+$ states: the extracted lifetime of the $2_1^+$ state is in perfect agreement with the literature, supporting the validity of the experimental method. 
For $^{106}$Sn the described TKEL-gate procedure was not required and, because of the presence of the long-lived isomer, the decay cascade of the $6_1^+$, $4_1^+$ and $2_1^+$ states have been taken into account while measuring the lifetime via DCM. 
Therefore, thanks to the powerful setup and the unconventional experimental technique, the lifetime of the $2_1^+$ and $4_1^+$ states has been measured, for the first time, in $^{106,108}$Sn. 
Table~\ref{tab:Results} summarizes the experimental results, showing the lifetimes and the derived reduced transition probabilities $B(E2)$ for $^{108}$Sn and $^{106}$Sn isotopes, as well as the theoretical values from the extension of the calculations of Ref.~\cite{banu2005sn}, obtained by employing the same interaction in the full $gds$ valence space for both protons and neutrons, using the effective charges $(e_\pi, e_\nu) = (1.35, 0.65)$, allowing up to $4p-4h$ excitations and without any seniority truncation.
The extracted $B(E2)$ values for the $^{106,108}$Sn isotopes are shown in Fig.~\ref{fig:B(E2)} together with all previous experimental results for the whole isotopic chain. 
The $B(E2; 2_1^+ \rightarrow 0_{g.s.}^+)$ strengths previously measured are compatible with the results obtained in this experiment, while for the $B(E2; 4_1^+ \rightarrow 2_1^+)$ values no data existed in this region. 
Unfortunately, the exoticity of $^{106}$Sn and the necessity to avoid the population of the states above the long-lived $6^+$ isomer result in a rather large statistical error on the $B(E2; 4_1^+ \to 2_1^+)$ value for this isotope.

\begin{table}[h]
\caption{\label{tab:Results} Measured lifetime of the excited states $I^\pi$ in $^{106,108}$Sn and corresponding $B(E2; I^\pi \to I^\pi-2)$ values. The last column shows the theoretical predictions from the extension of the calculations of Ref.~\cite{banu2005sn} (see text).}
\begin{ruledtabular}
\begin{tabular}{l c c c c c}
				&	\multirow{2}{*}{$I^\pi$}		&	\multirow{2}{*}{E$_\gamma$ [keV]}	&	\multirow{2}{*}{$\tau$ [ps]}	&	\multicolumn{2}{c}{B(E2) [$e^2 fm^4$]}  	\tstrut\bstrut \\
				&									&										&									&	Exp.				& 	Theo.  	\tstrut\bstrut \\
\hline
  $^{108}$Sn	&	$2_1^+$							&	1206.1								&	0.76 (8)						&	422 (44)			&	425		\tstrut\bstrut \\
				&	$4_1^+$							&	905.1								&	3.7 (2)							&	364 (20)			&	349		\tstrut\bstrut \\
  \hline
  $^{106}$Sn	&	$2_1^+$							&	1207.7								&	5 (4)							&	245 (132)			&	339		\tstrut\bstrut \\
				&	$4_1^+$							&	811.9								&	1.3 (7)							&	446 (334)			&	379		\tstrut\bstrut \\
\end{tabular}
\end{ruledtabular}
\end{table}

In view of the present experimental results, the interpretation of the data in the neutron-deficient tin isotopes were performed within the theoretical companion contribution by Zuker~\cite{CdSn}, this time taking into account also the newly measured $B(E2; 4^+_1 \rightarrow 2^+_1)$ values. 
Given a proper interaction it is very simple to describe the ``anomalous'' $B(E2;2_1^+ \to 0_{g.s.}^+)$ pattern of the Sn isotopes while the---hitherto unexplored---$B(E2;4_1^+ \to 2_1^+)$ behavior demands more care.

The interaction must be extracted from a realistic potential, properly renormalized and monopole corrected. 
All realistic potentials give very similar results~\cite{rmp}, N3LO was chosen~\cite{N3LOa} and $\vlk$-regularized~\cite{vlk}. 
The CDB~\cite{CDB} or AV18~\cite{AV18} potentials would yield the same results. 
The indispensable renormalizations amount to a---rigorously established---30\% boost of the quadrupole force and a phenomenological 40\% increase of the pairing force~\cite{mdz}. 
The monopole corrections were done by making the interaction monopole-free and adding the single particle spectrum of \A{101}Sn given in GEMO~\cite{gemo} (in parenthesis energies in MeV):

$g_{9/2}(-6.0),\,d_{5/2}(0.0),\,g_{7/2}(0.5),\, s_{1/2}(0.8),\, d_{3/2}(1.6)$ .

The resulting interaction is called I.3.4. 
Possible uncertainties concern the pairing content, so we have also tried I.3.0 and I.3.2 (no and 20\% corrections, respectively). 
Furthermore, according to the study of Ref.~\cite[Fig. 3.2.1]{Sn100}, an alternative to the GEMO spectrum (DZ in that Figure) is an extrapolated estimate (EX) equivalent to pushing up the $s_{1/2}$ orbital to 1.6 MeV, which results in the I.3.4s1.6 interaction.

The calculations were performed with the ANTOINE program~\cite{rmp} in $utM$ spaces in the $gds$ shell of up to $u\, g_{9/2}$-proton holes and $t\, g_{9/2}$-neutron holes, for a total of $M$ holes. 
Here we present $utM=202$ cases of m-scheme dimensions of up to $6 \cdot 10^7$, but it was checked that they reproduce well the $10^{10}$-dimensional $utM=444$ cases. 

\begin{figure}[t] 
\includegraphics[width=0.48\textwidth]{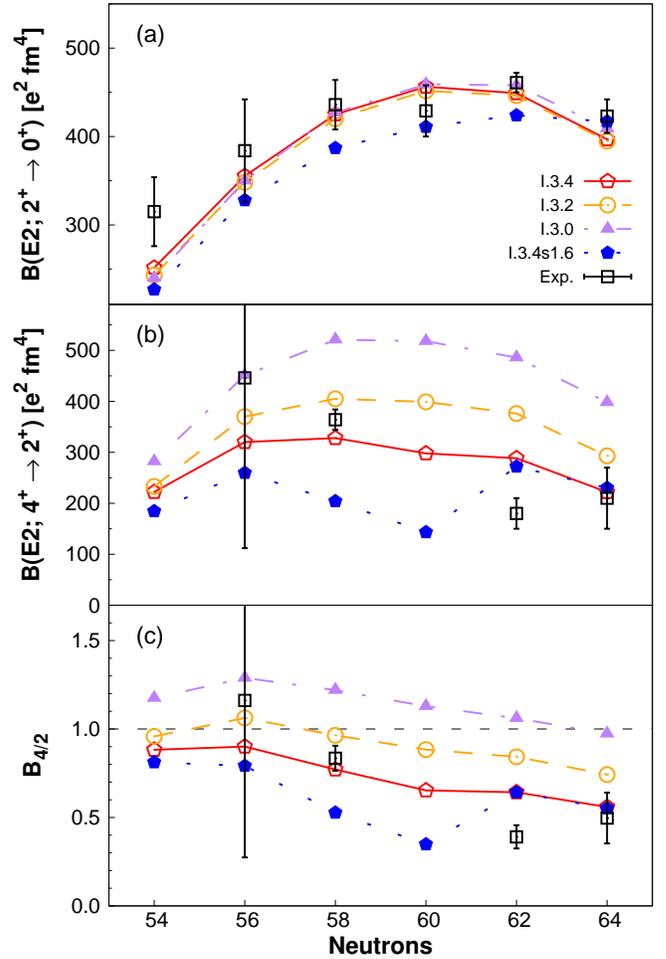}
\caption{\label{fig:be2042} (Color online) Experimental and calculated (a) $B(E2;2_1^+ \to 0_{g.s.}^+)$ and (b) $B(E2;4_1^+ \to 2_1^+)$ values and (c) $B_{4/2} \equiv B(E2;4_1^+ \to 2_1^+)/B(E2;2_1^+ \to 0_{g.s.}^+)$ ratio for Sn isotopes. Effective charges are $(e_\pi, e_\nu) = (1.40, 0.72)$ throughout. For $A=112$ and $114$ the neutron effective charge $e_\nu$ should be increased to 0.75 to account for the omission of the $h_{11/2}$ shell, that plays a small but significant role. 
Experimental $B(E2;2_1^+ \to 0_{g.s.}^+)$ values are the weighted averages of the Fig.~\ref{fig:B(E2)} results, while the $B(E2;4_1^+ \to 2_1^+)$ data comes from Ref.~\cite{jonsson1981collective} and the present work. I.3.0s1.6 has the $s_{1/2}$ single particle energy moved up by 800 keV with respect to GEMO.}
\end{figure}

Figure~\ref{fig:be2042} (a) establishes that the amount of pairing makes no difference in the $B(E2;2_1^+ \to 0_{g.s.}^+)$ transition probabilities. 
The shift in the position of the $s_{1/2}$ orbit has an influence, albeit minor. 
This behavior may explain why the pattern in the calculations of Togashi~\cite{PhysRevLett.121.062501} is identical to ours, while the wavefunctions are radically different~\cite{CdSn} in that they exhibit strong spin and mass dependence in the $g_{9/2}$ proton-hole occupancies, which are nearly constant in our case.

In the Fig.~\ref{fig:be2042} (b) both the pairing and the single particle shift make an enormous difference in the $B(E2;4_1^+ \to 2_1^+)$ behavior.

The context of these results is provided by the Pseudo-SU(3) symmetry which acts in the space of $sdg$ orbits above (and except) $g_{9/2}$.
For a strong enough quadrupole force and nearly degenerate single-particle spectrum, the system exhibits rotational features. 
In particular $B_{4/2} \equiv B(E2;4_1^+ \to 2_1^+)/B(E2;2_1^+ \to 0_{g.s.}^+) \approx 1.43$ (Alaga rule). 
All this is explained in~\cite{CdSn}.

In the case of strong quadrupole dominance, the $B(E2;2_1^+ \to 0_{g.s.}^+)$ pattern would be as in Fig.~\ref{fig:be2042} (a) and the $B(E2;4_1^+ \to 2_1^+)$ one would be the same multiplied by 1.43. 
This is close to the I.3.0 case, though the quotient (Fig.~\ref{fig:be2042} (c)) is reduced to about 1.25. 
It is further reduced for I.3.2. 
In the two I.3.4 cases, the proportionality between patterns is completely lost, but both are close to the observed values in \A{112-114}Sn~\cite{jonsson1981collective}. 
Our new measure in \A{108}Sn breaks the ambiguity in favor of the I.3.4 chosen standard with GEMO spectrum, providing a potentially interesting suggestion about the spectrum of \A{101}Sn.

The $B_{4/2}<1$ anomaly had been observed in \A{114}Xe~\cite{deangelis2002coherent}, in \A{114}Te~\cite{moller2005e2} and, more recently, in \A{172}Pt~\cite{cederwall2018lifetime}, where it is stressed that no theoretical explanation is available, except through pairing dominance. 
The present calculations bring a change in that they account for observation and point to the underlying mechanism: it is clear that quadrupole dominance holds for the $J=0_1,\, 2_1$ states, but is strongly challenged for the $4_1$ state through mixing with a pairing dominated intruder. 
What is missing is a characterization of such an intruder. 
A novel situation that demands further study.

In summary, the unconventional use of multi-nucleon transfer reactions with a plunger device has allowed to measure lifetimes in the neutron-deficient $^{106,108}$Sn isotopes. 
The TKEL-gate technique was used to overcome the experimental limitations due to the presence of low-lying isomers, leading to the first $B(E2; 4_1^+ \to 2_1^+)$ measurement in this mass region. 
The theoretical results show that the experimental trend of the $B(E2; 2_1^+ \to 0_{g.s.}^+)$ values in the mass region $104 \leq A \leq 114$ can be reproduced within the $gds$ model space. 
Traditionally, pairing was thought to be dominant in the Sn isotopes but the calculations indicate that the rather constant pattern of the $B(E2; 2_1^+ \to 0_{g.s.}^+)$ values is associated to quadrupole dominance, independent of the pairing strength. 
However, in the case of $B(E2; 4_1^+ \to 2_1^+)$ values pairing takes its revenge by becoming crucial.
What is beyond doubt is the importance of the very precise measurements in $^{108}$Sn, which have shown to open new perspectives in the understanding of the quadrupole-pairing interplay.

\vspace{5mm}

\input PRL_acknowledgement.tex   
\bibliography{Siciliano_108Sn,Zuker} 

\end{document}

%% file: PRL_AuthorList.tex
%
\author{M.~Siciliano}
	\affiliation{INFN, Laboratori Nazionali di Legnaro, Legnaro, Italy.}
	\affiliation{Dipartimento di Fisica e Astronomia, Universit\`a di Padova, Padua, Italy.}
	\affiliation{Irfu/CEA, Universit\'e de Paris-Saclay, Gif-sur-Yvette, France.}
\author{J.J.~Valiente-Dob\'on}
	\affiliation{INFN, Laboratori Nazionali di Legnaro, Legnaro, Italy.}
\author{A.~Goasduff}
	\affiliation{INFN, Laboratori Nazionali di Legnaro, Legnaro, Italy.}
	\affiliation{Dipartimento di Fisica e Astronomia, Universit\`a di Padova, Padua, Italy.}
	\affiliation{INFN, Sezione di Padova, Padua, Italy.}
\author{F.~Nowacki}
	\affiliation{IPHC, CNRS/IN2P3 Universit\'e de Strasbourg, Strasbourg, France.}
\author{A.P.~Zuker}
	\affiliation{IPHC, CNRS/IN2P3 Universit\'e de Strasbourg, Strasbourg, France.}
\author{D.~Bazzacco}
	\affiliation{INFN, Sezione di Padova, Padua, Italy.}
\author{A.~Lopez-Martens}
	\affiliation{CSNSM, CNRS/IN2P3, Universit\'e de Paris-Saclay, Orsay, France.}
\author{E.~Cl\'ement}
	\affiliation{Grand Acc\'el\'erateur National d’Ions Lourds, Irfu/CEA/DRF and CNRS/IN2P3, Caen, France.} 
\author{G.~Benzoni}
	\affiliation{INFN, Sezione di Milano, Milan, Italy.}
\author{T.~Braunroth}
	\affiliation{Institut f\"{u}r Kernphysik, Universit\"{a}t zu K\"{o}ln, Cologne, Germany.}
\author{N.~Cieplicka-Ory\'nczak}
	\affiliation{INFN, Sezione di Milano, Milan, Italy.}
	\affiliation{Instytut Fizyki J\c{a}drowej im. Henryka Niewodnicza\'nskiego, Polska Akademia Nauk, Krakow, Poland.}
\author{F.C.L.~Crespi}
	\affiliation{INFN, Sezione di Milano, Milan, Italy.}
	\affiliation{Dipartimento di Fisica, Universit\`a di Milano, Milan, Italy.}
\author{G.~de~France}
	\affiliation{Grand Acc\'el\'erateur National d’Ions Lourds, Irfu/CEA/DRF and CNRS/IN2P3, Caen, France.} 
\author{M.~Doncel}
	\affiliation{Universidad de Salamanca, Salamanca, Spain.}
\author{S.~Ert\"urk}
	\affiliation{\"Omer Halisdemir \"Universitesi, Ni\u{g}de, Turkey.}
\author{C.~Fransen}
	\affiliation{Institut f\"{u}r Kernphysik, Universit\"{a}t zu K\"{o}ln, Cologne, Germany.}
\author{A.~Gadea}
	\affiliation{Instituto de F\'isica Corpuscular, CSIC-Universidad de Valencia, Valencia, Spain.}
\author{G.~Georgiev}
	\affiliation{CSNSM, CNRS/IN2P3, Universit\'e de Paris-Saclay, Orsay, France.}
\author{A.~Goldkuhle}
	\affiliation{Institut f\"{u}r Kernphysik, Universit\"{a}t zu K\"{o}ln, Cologne, Germany.}
\author{U.~Jakobsson}
	\affiliation{Kungliga Tekniska H\"ogskolan, Stockholm, Sweden.}
\author{G.~Jaworski}
	\affiliation{INFN, Laboratori Nazionali di Legnaro, Legnaro, Italy.}
	\affiliation{\'Srodowiskowe Laboratorium Ci\c{e}\.zkich Jon\'ow, Uniwersytet Warszawski, Warsaw, Poland.}
\author{P.R.~John}
	\affiliation{Dipartimento di Fisica e Astronomia, Universit\`a di Padova, Padua, Italy.}
	\affiliation{INFN, Sezione di Padova, Padua, Italy.}
	\affiliation{Institut f\"ur Kernphysik, Technische Universit\"at Darmstadt, Darmstadt, Germany.}
\author{I.~Kuti}
	\affiliation{INR, Hungarian Academy of Sciences, Debrecen, Hungary.}
\author{A.~Lemasson}
	\affiliation{Grand Acc\'el\'erateur National d’Ions Lourds, Irfu/CEA/DRF and CNRS/IN2P3, Caen, France.} 
\author{H.~Li}
	\affiliation{Kungliga Tekniska H\"ogskolan, Stockholm, Sweden.}
\author{T.~Marchi}
	\affiliation{INFN, Laboratori Nazionali di Legnaro, Legnaro, Italy.}
\author{D.~Mengoni}
	\affiliation{Dipartimento di Fisica e Astronomia, Universit\`a di Padova, Padua, Italy.}
	\affiliation{INFN, Sezione di Padova, Padua, Italy.}
\author{C.~Michelagnoli}
	\affiliation{Grand Acc\'el\'erateur National d’Ions Lourds, Irfu/CEA/DRF and CNRS/IN2P3, Caen, France.} 
	\affiliation{Institut Laue-Langevin, Grenoble, France.}
\author{T.~Mijatovi\'c}
	\affiliation{Ru{d\llap{\raise 1.22ex\hbox{\vrule height 0.09ex width 0.4em}}\rlap{\raise 1.22ex\hbox{\vrule height 0.09ex width 0.04em}}}er Bo\v{s}kovi\'{c} Institute and University of Zagreb, Zagreb, Croatia.}
\author{C.~M\"uller-Gatermann}
	\affiliation{Institut f\"{u}r Kernphysik, Universit\"{a}t zu K\"{o}ln, Cologne, Germany.}
\author{D.R.~Napoli}
	\affiliation{INFN, Laboratori Nazionali di Legnaro, Legnaro, Italy.}
\author{J.~Nyberg}
	\affiliation{Institutionen f\"{o}r Fysik och Astronomi, K\"{a}rnfysik, Uppsala Universitet, Uppsala, Sweden.}
\author{M.~Palacz}
	\affiliation{\'Srodowiskowe Laboratorium Ci\c{e}\.zkich Jon\'ow, Uniwersytet Warszawski, Warsaw, Poland.}
\author{R.M.~P\'erez-Vidal}
	\affiliation{Instituto de F\'isica Corpuscular, CSIC-Universidad de Valencia, Valencia, Spain.}
\author{B.~Say\u{g}i}
	\affiliation{INFN, Laboratori Nazionali di Legnaro, Legnaro, Italy.}
	\affiliation{Ege \"Universitesi, \.Izmir, Turkey.}
\author{D.~Sohler}
	\affiliation{INR, Hungarian Academy of Sciences, Debrecen, Hungary.}
\author{S.~Szilner}
	\affiliation{Ru{d\llap{\raise 1.22ex\hbox{\vrule height 0.09ex width 0.4em}}\rlap{\raise 1.22ex\hbox{\vrule height 0.09ex width 0.04em}}}er Bo\v{s}kovi\'{c} Institute and University of Zagreb, Zagreb, Croatia.}
\author{D.~Testov}
	\affiliation{Dipartimento di Fisica e Astronomia, Universit\`a di Padova, Padua, Italy.}
	\affiliation{INFN, Sezione di Padova, Padua, Italy.}
\author{D.~Barrientos}
	\affiliation{CERN, Geneva, Switzerland.}
\author{B.~Birkenbach}
	\affiliation{Institut f\"{u}r Kernphysik, Universit\"{a}t zu K\"{o}ln, Cologne, Germany.}
\author{H.C.~Boston}
	\affiliation{Oliver Lodge Laboratory, University of Liverpool, Liverpool, UK.} 
\author{A.J.~Boston}
	\affiliation{Oliver Lodge Laboratory, University of Liverpool, Liverpool, UK.} 
\author{B.~Cederwall}
	\affiliation{Department of Physics, Royal Institute of Technology, Stockholm, Sweden.}
\author{D.M.~Cullen}
	\affiliation{Schuster Laboratory, University of Manchester, Manchester, UK.} 
\author{J.~Collado}
	\affiliation{Departamento de Ingenier\'ia Electr\'onica, Universitad de Valencia, Valencia, Spain.} 
\author{P.~D\'esesquelles}
	\affiliation{CSNSM, CNRS/IN2P3, Universit\'e de Paris-Saclay, Orsay, France.}
\author{J.~Dudouet}
	\affiliation{CSNSM, CNRS/IN2P3, Universit\'e de Paris-Saclay, Orsay, France.}
	\affiliation{IPN-Lyon, CNRS/IN2P3, Universit\'e de Lyon, Villeurbanne, France.} 
\author{C.~Domingo-Pardo}
	\affiliation{Instituto de F\'isica Corpuscular, CSIC-Universidad de Valencia, Valencia, Spain.}
\author{J.~Eberth}
	\affiliation{Institut f\"{u}r Kernphysik, Universit\"{a}t zu K\"{o}ln, Cologne, Germany.}
\author{F.J.~Egea-Canet}
	\affiliation{INFN, Laboratori Nazionali di Legnaro, Legnaro, Italy.}
\author{V.~Gonz\'alez}
	\affiliation{Departamento de Ingenier\'ia Electr\'onica, Universitad de Valencia, Valencia, Spain.} 
\author{D.S.~Judson}
	\affiliation{Oliver Lodge Laboratory, University of Liverpool, Liverpool, UK.}
\author{L.J.~Harkness-Brennan}
	\affiliation{Oliver Lodge Laboratory, University of Liverpool, Liverpool, UK.} 
\author{H.~Hess}
	\affiliation{Institut f\"{u}r Kernphysik, Universit\"{a}t zu K\"{o}ln, Cologne, Germany.}
\author{A.~Jungclaus}
	\affiliation{Instituto de Estructura de la Materia, CSIC, Madrid, Spain.} 
\author{W.~Korten}
	\affiliation{Irfu/CEA, Universit\'e de Paris-Saclay, Gif-sur-Yvette, France.}
\author{M.~Labiche}
	\affiliation{STFC Daresbury Laboratory, Warrington, UK.} 	
\author{A.~Lefevre}
	\affiliation{Grand Acc\'el\'erateur National d’Ions Lourds, Irfu/CEA/DRF and CNRS/IN2P3, Caen, France.} 
\author{S.~Leoni}
	\affiliation{INFN, Sezione di Milano, Milan, Italy.}
	\affiliation{Dipartimento di Fisica, Universit\`a di Milano, Milan, Italy.}
\author{A.~Maj}
	\affiliation{Instytut Fizyki J\c{a}drowej im. Henryka Niewodnicza\'nskiego, Polska Akademia Nauk, Krakow, Poland.}
\author{R.~Menegazzo}
	\affiliation{INFN, Sezione di Padova, Padua, Italy.}
\author{B.~Million}
	\affiliation{INFN, Sezione di Milano, Milan, Italy.}
\author{A.~Pullia}
	\affiliation{INFN, Sezione di Milano, Milan, Italy.} 
	\affiliation{Dipartimento di Fisica, Universit\`a di Milano, Milan, Italy.}
\author{F.~Recchia}
	\affiliation{Dipartimento di Fisica e Astronomia, Universit\`a di Padova, Padua, Italy.}
	\affiliation{INFN, Sezione di Padova, Padua, Italy.}
\author{P.~Reiter}
	\affiliation{Institut f\"{u}r Kernphysik, Universit\"{a}t zu K\"{o}ln, Cologne, Germany.}
\author{M.D.~Salsac}
	\affiliation{Irfu/CEA, Universit\'e de Paris-Saclay, Gif-sur-Yvette, France.} 
\author{E.~Sanchis}
	\affiliation{Departamento de Ingenier\'ia Electr\'onica, Universitad de Valencia, Valencia, Spain.} 
\author{O.~Stezowski}
	\affiliation{IPN-Lyon, CNRS/IN2P3, Universit\'e de Lyon, Villeurbanne, France.} 
\author{Ch.~Theisen}
	\affiliation{Irfu/CEA, Universit\'e de Paris-Saclay, Gif-sur-Yvette, France.} 
\author{M.~Zieli\'nska}
	\affiliation{Irfu/CEA, Universit\'e de Paris-Saclay, Gif-sur-Yvette, France.}

\vskip 0.25cm

%% file: PRL_acknowledgement.tex
The authors would like to thank the AGATA and VAMOS collaborations. 
Special thanks go to the GANIL technical staff for their help in setting up the apparatuses and the good quality beam and to A. Poves for his fruitful comments. 
This work was partially supported by the European Union’s Seventh Framework Programme for Research and Technological Development (grant no. 262010). 
The work was also supported (J.N.) by Swedish Research Council, (B.S.) by the Scientific and Technological Council of Turkey (TUBITAK) under the project no. 114F473, (A.G., R.P. and A.J.) by the Ministerio de Econom\'ia y Competitividad and the Ministry of Science of Spain under the grant agreements nos. EEBB-I-15-09671, FPA2017-84756-C4 and SEV-2014-0398 and by the EU-FEDER funds, (T.M and S.S.) by the Croatian Science Foundation under the project no. 7194, (I.K and D.S.) by the Hungarian National Research and Innovation Office (NKFIH) under the project nos. K128947, PD124717 and GINOP-2.3.3-15-2016-00034, (M.P.) by the Polish National Science Centre with the grants nos. 2014/14/M/ST2/00738, 2016/22/M/ST2/00269 and 2017/25/B/ST2/01569 under the COPIN-IN2P3, COPIGAL and POLITA projects. 